\documentclass[aps,prl,twocolumn,floatfix,showpacs,11pt,reprint]{revtex4-1}

\usepackage{graphicx}
\usepackage{amsmath, amsfonts, amssymb, bm}

\usepackage{amstext}
\usepackage{mathrsfs}

\begin{document}
\title{Ultrarelativistic electron states in a general background electromagnetic field}
\author{A.~Di Piazza}
\email{dipiazza@mpi-hd.mpg.de}
\affiliation{Max-Planck-Institut f\"ur Kernphysik, Saupfercheckweg 1, D-69117 Heidelberg, Germany}

\begin{abstract}
The feasibility of obtaining exact analytical results in the realm of QED in the presence 
of a background electromagnetic field is almost exclusively limited to a few 
tractable cases, where the Dirac equation in the corresponding background field 
can be solved analytically. This circumstance has restricted, in particular, the theoretical analysis
of QED processes in intense laser fields to within the plane-wave approximation
even at those high intensities, achievable experimentally only by tightly focusing
the laser energy in space. Here, within the Wentzel-Kramers-Brillouin (WKB) or eikonal
approximation, we construct analytically single-particle electron 
states in the presence of a background electromagnetic field of general space-time structure
in the realistic assumption that the initial energy of the electron is the largest 
dynamical energy scale in the problem. The relatively compact expression of these states
opens, in  particular, the possibility of investigating analytically 
strong-field QED processes in the presence of spatially focused laser beams, 
which is of particular relevance in view of the upcoming experimental campaigns in this field.
\end{abstract}

\maketitle

The predictions of QED have been confirmed with outstanding
precision in numerous experiments. The impressive agreement between
the theoretical and the experimental value of the electron $(g-2)$-factor 
is customarily quoted as a prominent example \cite{Hanneke_2008}. 
However, the experimental scrutiny of QED becomes
much less thorough when processes are involved, occurring in the presence of
a strong background electromagnetic field, i.e. of the order of 
$F_{cr}=m^2c^3/\hbar |e|=1.3\times 10^{16}\;\text{V/cm}=4.4\times 10^{13}\;\text{G}$
(here $m$ and $e<0$ are the electron mass and charge, respectively) \cite{Landau_b_4_1982}.
The main reason is that these values largely exceed the field strengths available
in laboratories. An important exception 
is represented by the electric field of highly-charged ions (charge number 
$Z\sim 1/\alpha$, with $\alpha=e^2/\hbar c\approx 1/137$) 
at the typical QED length $\lambda_C=\hbar/mc=3.9\times 10^{-11}\;\text{cm}$ \cite{Mohr_1998,Baur_2007}. 
Indeed, numerous experiments on 
processes occurring in the presence of highly-charged 
ions \cite{Milstein_1994,Akhmadaliev_2002,Sturm_2011,Volotka_2013}
have already successfully confirmed the predictions of QED. Correspondingly,
advanced analytical methods \cite{Lee_2000}, have been developed 
to interpret accurate experimental data beyond the exactly-solvable 
Coulomb model of the ionic field.

Modern high-power lasers represent an alternative source
of intense electromagnetic fields structurally 
thoroughly different from atomic fields. 
Although the amplitude $F_0$ of the strongest
laser pulse ever produced is about $10^{-4}\,F_{cr}$ \cite{Yanovsky_2008},
it can be boosted to an effective strength $F_0^*\sim F_{cr}$ in the rest-frame of ultrarelativistic particles
colliding with the laser beam \cite{Landau_b_2_1975}. This 
principle has been exploited at SLAC to perform 
the so-far unique experimental campaign on strong-laser field QED \cite{SLAC},
employing a laser with photon energy $1\;\text{eV}$ and amplitude $F_0=2.7\times 10^{10}\;\text{V/cm}$, and an almost counter-propagating electron beam with 
energy of $45\;\text{GeV}$ ($F_0^*\approx 0.3\,F_{cr}$). The relatively large pulse spot 
area ($\sim 60\;\text{$\mu$m$^2$}$) allowed for the experimental results being well reproduced 
theoretically within the plane-wave field approximation.

Approximating the laser field as a plane wave allows one to solve
exactly the Dirac equation in the resulting background electromagnetic 
field \cite{Volkov_1935}. The corresponding electron single-particle states
(Volkov states) have been extensively employed to investigate 
different strong-field QED processes \cite{Nikishov_1964,Reiss_2009,Boca_2009,Mackenroth_2011,Seipt_2011,Krajewska_2012,Loetstedt_2009,Seipt_2012,Mackenroth_2013,Reiss_1962,Heinzl_2010,Titov_2012,Mueller_2011,Meuren_2011,Hu_2010,Ilderton_2011}
(see also the recent reviews \cite{Ehlotzky_2009,Ruffini_2010,Di_Piazza_2012}). 
Correspondingly, Particle in Cell (PIC) codes including strong-field QED effects \cite{Elkina_2011,Nerush_2011,Ridgers_2012} in the dynamics of laser-irradiated plasmas employ expressions of the QED rates 
calculated in the plane-wave (local-constant-crossed-field) approximation. However, 
no analytical calculations in strong-field QED have been performed so far, which
also include self-consistently the spatial focusing of the
laser beam. This is especially desirable as ultra-high intensities 
are attained nowadays by spatially focusing the laser energy almost down 
to the diffraction limit.

In the present Letter we determine analytically the electron single-particle
states in the presence of a strong background electromagnetic field
of general space-time structure in the experimentally relevant case of an ultrarelativistic 
electron. In the realistic assumption that the initial 
energy of the electron is the largest dynamical energy scale in the problem,
we first determine the classical worldline of the electron
and then we construct the corresponding quantum states
in the Wentzel-Kramers-Brillouin (WKB) or eikonal approximation \cite{Landau_b_2_1975,Landau_b_3_1977}.
The availability of such single-particle states and, in particular,
their relatively compact expression open the possibility of investigating analytically 
and in a systematic way strong-field QED processes in the presence of 
intense background fields with complex space-time structure as, e.g., 
those of tightly-focused laser beams. 

To date, two methods have been developed to investigate QED 
processes in the presence of a virtually arbitrary background electromagnetic field.
However, the first one, based on the quasiclassical operator 
technique \cite{Baier_1967,Baier_1968}, allows one to obtain results only at the leading order in the quasiclassical, 
ultrarelativistic limit and does not contain a general prescription 
on how to calculate neither the amplitude of a generic QED process nor 
high-order corrections. The second one, instead, employs the so-called
``trajectory-coherent-states'' (TCS) \cite{Belov_1989,Bagrov_1993}, which are relativistic 
electron wave-functions localized near the classical electron's trajectory. However,
the expression of the TCS is extremely cumbersome and of limited
use for practical calculations.

We first consider the classical problem of an ultrarelativistic
electron moving in a background electromagnetic field, described by
the four-vector potential $A^{\mu}(x)$ in the Lorentz gauge $\partial_{\mu}A^{\mu}=0$ 
(here and below, units with $c=1$ are employed). We work in the laboratory
frame where the electron initial four-momentum is 
$p_0^{\mu}=(\varepsilon_0,\bm{p}_0)=(\sqrt{m^2+\bm{p}_0^2},\bm{p}_0)$ 
and we have in mind the case where the background electromagnetic field represents 
an intense, short, and tightly focused laser beam. Thus, we also assume that the field tensor $F^{\mu\nu}(x)=\partial^{\mu}A^{\nu}(x)-
\partial^{\nu}A^{\mu}(x)=(\bm{E}(x),\bm{B}(x))$ is localized in space and time, that it
has a maximum amplitude $F_0$, and that
it is characterized by a typical angular frequency $\omega_0$, 
such that the classical nonlinearity parameter $\xi_0=|e|F_0/m\omega_0$ (see \cite{Heinzl_2009} for a manifestly covariant and gauge-invariant definition of the parameter $\xi_0$ (see also \cite{Nikishov_1964})) satisfies the strong inequalities: $m\ll m\xi_0\ll\varepsilon_0$.
The above assumptions well fit 
present and near-future experimental conditions envisaged to test 
strong-field QED with intense lasers. In fact,
even next-generation of $10\;\text{PW}$ Ti:Sa lasers \cite{APOLLON_10P}
(central wavelength $\lambda_0=0.8\;\text{$\mu$m}$) are 
realistically expected not to exceed a peak intensity of $I_0\sim 10^{23}\;\text{W/cm$^2$}$, 
corresponding to $F_0\sim 6\times 10^{12}\;\text{V/cm}=2\times 10^{10}\;\text{G}$, $\xi_0\sim 160$, 
and $m\xi_0\sim 80\;\text{MeV}$. Such a field amplitude is effectively boosted 
to the critical value $F_{cr}$ in the rest frame of an electron with an energy of $\varepsilon_0\gtrsim 500\;\text{MeV}$,
which is about six times $m\xi_0$. In addition, electron beams with energies of about $2\;\text{GeV}$ have 
been already demonstrated experimentally also with laser-plasma 
accelerators \cite{Wang_2013}. Our starting point is the Lorentz equation: $dp^{\mu}/ds=(e/m)F^{\mu\nu}p_{\nu}$,
where $p^{\mu}=(\varepsilon,\bm{p})=(\sqrt{m^2+\bm{p}^2},\bm{p})$ is the electron four-momentum and $s$ is its proper time. According to the analytical solution of the Lorentz equation in a plane-wave \cite{Landau_b_2_1975},
the condition $m\xi_0\ll \varepsilon_0$ in the laboratory frame ensures that the electron
will be only slightly deflected from its initial direction by the background field
in the physically relevant situation where it is initially counterpropagating with respect
to the laser field. Thus, rather than working with manifestly covariant equations it is convenient to introduce the light-cone coordinates $\phi=t-\bm{n}\cdot \bm{x}$,
$\tau=(t+\bm{n}\cdot \bm{x})/2$, and $\bm{x}_{\perp}=\bm{x}-(\bm{n}\cdot \bm{x})\bm{n}$,
with $\bm{n}=\bm{p}_0/|\bm{p}_0|$, and the
quantities $n^{\mu}=(1,\bm{n})$, $\tilde{n}^{\mu}=(1/2)(1,-\bm{n})$, and $a_j^{\mu}=(0,\bm{a}_j)$,
with $j=1,2$ (in this respect, see also \cite{Kogut_1970}, where vacuum QED has been
formulated by employing light-cone coordinates in the so-called ``infinite-momentum frame''). The quantities $\bm{a}_1$ and $\bm{a}_2$ introduced above are two unit-vectors perpendicular
to $\bm{n}$ and to each other, and such that $\bm{a}_1\times \bm{a}_2=\bm{n}$.
An arbitrary four-vector $v^{\mu}=(v^0,\bm{v})$ can be expressed as:
$v^{\mu}=v_+n^{\mu}+v_-\tilde{n}^{\mu}+v_1a_1^{\mu}+v_2a_2^{\mu}$, where $v_+=(\tilde{n}v)=(v^0+\bm{n}\cdot\bm{v})/2$, 
$v_-=(nv)=v^0-\bm{n}\cdot\bm{v}$, and $v_j=-(a_jv)=\bm{a}_j\cdot\bm{v}$ (note that $a_j^2=-\bm{a}_j^2=-1$). In the original light-cone notation \cite{Dirac_1949}, the direction $\bm{n}$ was chosen as the ``third'' one, i.e., $\bm{n}=(0,0,1)$. However, for the sake of convenience in the use of the final results, we prefer to keep $\bm{n}$ as an arbitrary unit vector.

The on-shell condition $p^2=m^2$ implies that $p_-=(m^2+\bm{p}^2_{\perp})/2p_+$ and, in the physical situation of interest here, we require that the condition $|\bm{p}_{\perp}|\sim m\xi_0\ll p_+$ is satisfied in the laboratory frame, i.e., that the quantity $p_+\approx \varepsilon$ is the largest dynamical energy scale in the problem. By parametrizing the electron trajectory via the ``time'' $\tau$, the three independent components of the Lorentz equation can be written in the convenient form
\begin{align}
\label{L_+}
\frac{dp_+}{d\tau}&=eE_n+\frac{e}{2}\frac{\bm{F}_m\cdot\bm{p}_{\perp}}{p_+},\\
\label{L_j}
\frac{d\bm{p}_{\perp}}{d\tau}&=e\bm{F}_p-eB_n\frac{\bm{n}\times\bm{p}_{\perp}}{p_+}+\frac{e}{4}\frac{m^2+\bm{p}_{\perp}^2}{p_+^2}\bm{F}_m,
\end{align}
where the light-cone components of the field tensor have been expressed in terms of the electromagnetic field as $F_{\tilde{n},n}=\tilde{n}_{\mu}F^{\mu\nu}n_{\nu}=\bm{n}\cdot\bm{E}=E_n$, $F_{\tilde{n},j}=\tilde{n}_{\mu}F^{\mu\nu}a_{j,\nu}=\bm{a}_j\cdot\bm{F}_m/2$, $F_{n,j}=n_{\mu}F^{\mu\nu}a_{j,\nu}=\bm{a}_j\cdot\bm{F}_p$ and $F_{1,2}=a_{1,\mu}F^{\mu\nu}a_{2,\nu}=-\bm{n}\cdot\bm{B}=-B_n$, with $\bm{F}_{p/m}=\bm{E}_{\perp}\pm\bm{n}\times\bm{B}$. The idea now is to solve Eqs. (\ref{L_+})-(\ref{L_j}) iteratively by exploiting the appearance of different powers of the small quantity $|\bm{p}_{\perp}|/p_+$. We assume that the light-cone components of $F^{\mu\nu}$ have all the same order of magnitude and that the relative size of each term is determined by the power of the quantity $1/p_+$. If this is not the case, in fact, a careful analysis is required, as the mentioned hierarchy could be altered. This is expected to occur more likely in the idealized case of highly symmetric fields. For example, for a constant and uniform magnetic field perpendicular to $\bm{n}$, it is $B_n\equiv 0$ and the term proportional to $1/p_+$ in Eq. (\ref{L_j}) vanishes, unlike the one proportional to $1/p^2_+$. We also note that for a tightly focused Gaussian beam counterpropagating with respect to the electron, it is $|E_n|\sim 0.1|\bm{F}_p|$ \cite{Salamin_2002,Narozhny_2004}.

Now, the field components in Eqs. (\ref{L_+})-(\ref{L_j}) are calculated along the electron's trajectory. Since $p^{\mu}=p_+dx^{\mu}/d\tau$, the following exact equations for the electron's ``spatial'' coordinates $\bm{r}(\tau)=(\phi(\tau),\bm{x}_{\perp}(\tau))$ as functions of $\tau$ can be derived:
\begin{align}
\label{L_phi}
\frac{d\phi}{d\tau}=&\frac{m^2+\bm{p}_{\perp}^2}{2p_+^2},\\
\label{L_x_j}
\begin{split}
\frac{d^2\bm{x}_{\perp}}{d\tau^2}=&e\frac{\bm{F}_p}{p_+}-eE_n\frac{\bm{p}_{\perp}}{p_+^2}-eB_n\frac{\bm{n}\times\bm{p}_{\perp}}{p_+^2}\\
&+\frac{e}{4}\frac{m^2+\bm{p}^2_{\perp}}{p^3_+}\bm{F}_m-\frac{e}{2}(\bm{F}_m\cdot\bm{p}_{\perp})\frac{\bm{p}_{\perp}}{p_+^3}.
\end{split}
\end{align}
We set the initial conditions at a given time $\tau_0=(t_0+\bm{n}\cdot \bm{x}_0)/2$ as $\bm{r}(\tau_0)=\bm{r}_0=(\phi_0,\bm{x}_{0,\perp})$, with $\phi_0=t_0-\bm{n}\cdot \bm{x}_0$, and as $p_+(\tau_0)=p_{0,+}$ (recall that, by definition, $\bm{p}_{\perp}(\tau_0)=\bm{p}_{0,\perp}=\bm{0}$ and that the on-shell condition implies that $p_-(\tau_0)=p_{0,-}=m^2/2p_{0,+}$). We also assume that $A^{\mu}(\tau_0,\bm{r})=0$, with $\bm{r}=(\phi,\bm{x}_{\perp})$. 

By denoting the quantities calculated up to terms proportional to $1/p_{0,+}$ via the upper index $(1)$, we have that
\begin{equation}
\label{In_cond}
\bm{r}^{(1)}(\tau)=\left(\phi_0,\bm{x}_{0,\perp}+\frac{e}{p_{0,+}}\int_{\tau_0}^{\tau}d\tau'\bm{G}_p(\tau',\bm{r}_0)\right),
\end{equation}
where $\bm{G}_p(\tau,\bm{r}_0)=\int_{\tau_0}^{\tau}d\tau'\bm{F}_p(\tau',\bm{r}_0)$.

By substituting the expression of $\bm{r}^{(1)}(\tau)$ in Eqs. (\ref{L_+})-(\ref{L_j}) and by integrating them, we obtain that
\begin{align}
\label{u_+^1}
\begin{split}
p^{(1)}_+(\tau)=&p_{0,+}+\int_{\tau_0}^{\tau}d\tau'\bigg[eE_n(\tau',\bm{r}_0)\\
&+\frac{e^2}{p_{0,+}}\bm{G}_p(\tau',\bm{r}_0)\cdot\bm{\nabla}_{\perp}\int_{\tau'}^{\tau} d\tau''E_n(\tau'',\bm{r}_0)\\
&+\frac{e^2}{2p_{0,+}}\bm{F}_m(\tau',\bm{r}_0)\cdot\bm{G}_p(\tau',\bm{r}_0)\bigg],
\end{split}\\
\label{u_j^1}
\begin{split}
\bm{p}^{(1)}_{\perp}(\tau)=&\int_{\tau_0}^{\tau}d\tau'\bigg[e\bm{F}_p(\tau',\bm{r}_0)\\
&+\frac{e^2}{p_{0,+}}\bm{G}_p(\tau',\bm{r}_0)\cdot\bm{\nabla}_{\perp} \int_{\tau'}^{\tau} d\tau''\bm{F}_p(\tau'',\bm{r}_0)\\
&-\frac{e^2}{p_{0,+}}B_n(\tau',\bm{r}_0)(\bm{n}\times\bm{G}_p(\tau',\bm{r}_0))\bigg],
\end{split}\\
\label{u_-^1}
p^{(1)}_-(\tau)=&\frac{m^2+e^2\bm{G}^2_p(\tau,\bm{r}_0)}{2p_{0,+}}.
\end{align}
The condition $|\bm{p}^{(1)}_{\perp}(\tau)|\ll p^{(1)}_+(\tau)$ in the laboratory frame ensures that our approximated solution is accurate (see Eqs. (\ref{L_+})-(\ref{L_j})) and it is fulfilled if $|e\bm{G}_p(\tau,\bm{r}_0)|\ll p_{0,+}$. Since tightly focused laser pulses are usually localized in a space (time) region of the order of a few laser central wavelengths (periods), the above condition is equivalent in order of magnitude to the requirement $m\xi_0\ll \varepsilon_0$ in the relevant case of an electron initially counterpropagating with respect to the laser beam. In order to highlight the qualitative novelties in the theoretical predictions brought about by the inclusion of the laser spatial focusing, in Fig. 1 we plot the momentum $p_z$ in units of $m$ as a function of the quantity $\omega_0 t$ for an electron (initial conditions $\bm{x}_0=(0,0,1.5\;\text{$\mu$m})$ and $\bm{p}_0=(0,-260\;\text{MeV},0)$ at $t_0=0$) initially counterpropagating with respect to a Ti:Sa, Gaussian, $\sin^2$-pulse beam \cite{Salamin_2002}, linearly polarized along the $z$-direction with duration $T=16\;\text{fs}$, spot radius $w_0=1.5\;\text{$\mu$m}$, and peak intensity $I_0=5.7\times 10^{22}\;\text{W/cm$^2$}$ ($\xi_0=110$).
\begin{figure}
\begin{center}
\includegraphics[width=0.8\columnwidth]{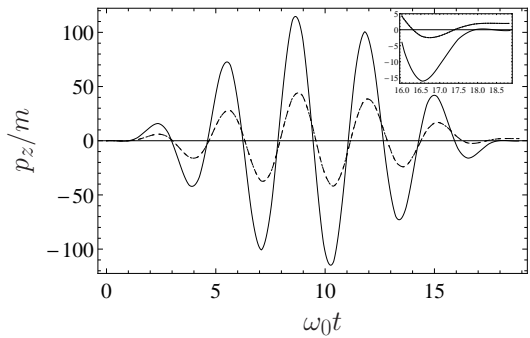}
\caption{Electron transverse momentum $p_z$ in units of the electron mass as a function of $\omega_0t$ for numerical values and details given in the text.}
\end{center}
\end{figure}
The continuous (dashed) line indicates the results of the numerical integration of the Lorentz equation neglecting (including) the beam spatial focusing. The dotted line, on top of the dashed one, corresponds to the analytical result from Eq. (\ref{u_j^1}) (discrepancies between the numerical and the analytical results arise at the third significant digit). The final value of $p_z$ in the case of the Gaussian beam is $1\;\text{MeV}$, whereas it vanishes within the plane-wave approximation (see the inset in Fig. 1), according to the Lawson-Woodward theorem (see, e.g., \cite{Troha_1999}). Note that the corresponding divergence of $4\;\text{mrad}$ is larger than already demonstrated electron beam divergences (e.g., of $0.45\;\text{mrad}$ at a beam energy of $245\;\text{MeV}$ \cite{Weingartner_2012}). In general, by setting $\tau\to\infty$ in Eqs. (\ref{u_+^1})-(\ref{u_-^1}), the relation between the final four-momentum $p^{(1),\mu}(\infty)$ and the initial one $p_0^{\mu}$ can be obtained. Another qualitatively new feature brought about by the laser focusing in the above-mentioned physical setup is that the quantity $p_+(\tau)$ is no more a constant of motion as in the plane-wave case and, indeed, the correction to the plane-wave result depends on the longitudinal electric field of the laser (see Eq. (\ref{u_+^1})).

It is interesting to observe that by keeping only the leading-order term of each component of the four-momentum obtained above (i.e., by approximating $p_+(\tau)\approx p_{0,+}$, $\bm{p}_{\perp}(\tau)\approx e\bm{G}_p(\tau,\bm{r}_0)$, and $p_-(\tau)\approx [m^2+e^2\bm{G}^2_p(\tau,\bm{r}_0)]/2p_{0,+}$), the corresponding expression coincides with the exact four-momentum of an electron in a background plane-wave like field depending on $\tau$ and calculated at the initial coordinates $\bm{r}_0$. This is in agreement with the classical result that an ultrarelativistic particle ``sees'' an arbitrary background field in its rest frame as a plane-wave-like field at leading order \cite{Landau_b_2_1975}. Note that, although it might be more convenient to express the four-momentum obtained above in terms of the four-potential $A^{\mu}(x)$ as in the plane-wave case, we prefer to employ the physical observable electromagnetic field.

We pass now to the quantum case and we consider the Dirac equation $[\gamma^{\mu}(i\hbar\partial_{\mu}-eA_{\mu})-m]\psi=0$, where $\gamma^{\mu}$ are the Dirac matrices and $\psi(x)$ is the bi-spinor electron wave function \cite{Landau_b_4_1982}. Based on the general argument that the De Broglie length of an ultrarelativistic particle is very small, we apply the WKB method \cite{Landau_b_3_1977} and look for a solution of the form $\psi(x)=\exp[iS(x)/\hbar]\varphi(x)$, where $S(x)$ turns out to be the classical electron action \cite{Pauli_1932,Rubinow_1963,Maslov_1981}. For an electron with initial four-momentum $p_0^{\mu}$ and spin quantum number $\sigma_0$, the positive-energy wave function $\psi^{(1)}_{p_0,\sigma_0}(\tau,\bm{r})$ up to the first order in $1/p_{0,+}$ reads (see the Supplemental Material (SM) for a detailed derivation):
\begin{equation}
\label{psi_0}
\begin{split}
\psi^{(1)}_{p_0,\sigma_0}(\tau,\bm{r})=&e^{iS^{(1)}_{p_0}(\tau,\bm{r})/\hbar}\bigg[1-\frac{e}{2}\int_{\tau_0}^{\tau}\frac{d\tau'}{p_{0,+}}\{\bm{\nabla}_{\perp}\cdot\bm{G}_p(\tau',\bm{r})\\
&-i\bm{\Sigma}\cdot[\bm{B}(\tau',\bm{r})-\bm{n}\times\bm{E}(\tau',\bm{r})]\}\bigg] \frac{u_{p_0,\sigma_0}}{\sqrt{2\varepsilon_0}},
\end{split}
\end{equation}
where
\begin{equation}
\label{S^1}
\begin{split}
S^{(1)}_{p_0}(\tau,\bm{r})=&-p_{0,+}\phi-\frac{m^2}{2p_{0,+}}\tau\\
&-\int_{\tau_0}^{\tau}d\tau'\bigg[eA_-(\tau',\bm{r})+\frac{e^2}{2p_{0,+}}\bm{G}^2_p(\tau',\bm{r})\bigg],
\end{split}
\end{equation}
where $\bm{\Sigma}=-i\gamma^1\gamma^2\gamma^3\bm{\gamma}$, where $u_{p_0,\sigma_0}$ is the usual constant free bi-spinor  \cite{Landau_b_4_1982}, and where a unity quantization volume is assumed. In the SM it is shown that in the relevant case of a strong ($\xi_0\gg 1$), tightly-focused ($w_0\approx \lambda_0$), and short ($T\sim \lambda_0$) optical ($\lambda_0\sim 1\;\text{$\mu$m}$) laser field the only restrictive condition for the validity of the wave function $\psi^{(1)}_{p_0,\sigma_0}(\tau,\bm{r})$ is the classical one $m\xi_0\ll\varepsilon_0$. The wave function $\psi^{(1)}_{p_0,\sigma_0}(\tau,\bm{r})$ reduces to the one obtained in \cite{Akhiezer_1979} in the particular case of a background time-independent scalar potential (see also \cite{Blankenbecler_1987}). Also, ultrarelativistic wave functions for scalar particles \cite{Blackenbecler_1970} and two-particles scattering amplitudes \cite{Cheng_1969,Abarbanel_1969} have been derived in the context of high-energy scattering in QED in the leading-order eikonal approximation, which corresponds in our notation to neglect terms proportional to $1/p_{0,+}$. However, keeping these terms is essential here, e.g., to recover the plane-wave results. In fact, if the background field is a plane-wave field depending on $\tau$, the state in Eq. (\ref{psi_0}) coincides within our approximations with the corresponding Volkov state \cite{Landau_b_4_1982}. In this respect, we note that the average spin $\hbar\bm{\zeta}$ (see the discussion below Eq. (11) in the SM) in a Volkov state describing an electron initially counterpropagating with respect to a linearly polarized plane wave never acquires a component along the magnetic field of the plane wave if it is initially along the electron momentum \cite{Landau_b_4_1982}. Whereas, by employing the wave function $\psi^{(1)}_{p_0,\sigma_0}(\tau,\bm{r})$, this does occur in the case of a focused laser field. In the particular setup mentioned below Eq. (\ref{u_-^1}), the average spin acquires an $x$-component $\hbar\zeta^{(1)}_x(\tau,\bm{r})=(e\hbar/p_{0,+})\int_{\tau_0}^{\tau}d\tau'[E_x(\tau',\bm{r})-B_z(\tau',\bm{r})]$, which vanishes identically in the corresponding plane-wave case.

The negative-energy electron states $\psi^{(1)}_{-p_0,-\sigma_0}(\tau,\bm{r})$ can be obtained via the substitutions $p_0^{\mu}\to-p_0^{\mu}$ and $\sigma_0\to-\sigma_0$ in Eq. (\ref{psi_0}) except that in $1/\sqrt{2\varepsilon_0}$, with the resulting quantity $u_{-p_0,-\sigma_0}$ being the free negative-energy constant bi-spinor \cite{Landau_b_4_1982}. In-/out-states are obtained by performing the limit $\tau_0\to\mp\infty$ in Eq. (\ref{psi_0}) and in the action $S_{p_0}(\tau,\bm{r})$, with the quantum numbers $p_0$ and $\sigma_0$ corresponding to the asymptotic four-momentum and spin outside the field at $\tau_0\to\mp\infty$. 

Once the single-particle positive- and negative-energy, in- and -out-states $\psi^{(\text{in/out})}_{\pm p,\pm\sigma}(x)$ in ordinary coordinates have been determined (the upper index $(1)$ from the single-particle states has been removed for the sake of notational simplicity), the matrix element $M_{f,i}$ of a typical process as nonlinear Compton scattering can be calculated as (see, e.g., Eq. (4.1.32) in \cite{Fradkin_1991})
\begin{equation}
M_{f,i}=-ie\sqrt{\frac{2\pi}{\omega}}\int d^4x\,\bar{\psi}^{(\text{out})}_{p_f,\sigma_f}(x)\hat{e}^*_{k,\lambda}\psi^{(\text{in})}_{p_i,\sigma_i}(x)e^{i(kx)}.
\end{equation}
Here, the quantities $p_{i/f}$ and $\sigma_{i/f}$ characterize the initial/final electron, whereas the emitted photon has four-momentum $k^{\mu}=(\omega,\bm{k})$ and polarization four-vector $(e_{k,\lambda})^{\mu}$ ($\hat{e}^*_{k,\lambda}=\gamma^{\mu}(e^*_{k,\lambda})_{\mu}$). A semi-quantitative analysis of the matrix element $M_{f,i}$ already reveals new features in the focused-field case with respect to the plane-wave one  (and also to the locally constant-crossed-field one, which is relevant for PIC codes). First, unlike that in the plane-wave case, we can introduce here the concept of transverse formation region(s) of radiation with respect to the laser propagation direction, analogous to the concept of ``impact parameter'' in, e.g., electron-nucleus collision \cite{Akhiezer_1979}. In the quasiclassical limit, this can be physically understood as, unlike that in a plane wave, electron trajectories in a focused field differing only by the initial transverse position contribute in general with different phases to the radiation process. Now, Eqs. (\ref{psi_0})-(\ref{S^1}) show that $\psi^{(\text{in/out})}_{p_{i/f},\sigma_{i/f}}(x)\sim\exp[iS^{(\text{in/out})}_{p_{i/f}}(x)/\hbar]u_{p_{i/f},\sigma_{i/f}}$, with $\lim_{t\to\mp\infty}S^{(\text{in/out})}_{p_{i/f}}(x)=-(p_{i/f}x)$. In the quasiclassical, ultrarelativistic regime at $\xi_0\gg 1$, the matrix element $M_{f,i}$ can be evaluated approximately via the saddle-point method (see, e.g., \cite{Di_Piazza_2012}). For any saddle point $x_l$ characterized by the conditions $\Delta \pi^{\mu}(x_l)=\pi^{(\text{out})\mu}_f(x_l)+\hbar k^{\mu}-\pi^{(\text{in})\mu}_i(x_l)=0$, with $\pi^{(\text{in/out})\mu}_{i/f}(x)=-\partial^{\mu}[S^{(\text{in/out})}_{p_{i/f}}(x)]$, one can estimate the transverse formation regions $l_s$, with $s=\{x,z\}$, from the resulting quadratic term in $(x_s-x_{l,s})^2$ in the exponent as $l_s\sim \sqrt{\hbar/\vert\partial \Delta \pi_s/\partial x_s\vert}$, where all quantities are calculated at $x_l$. Moreover, contrary to a plane wave, a focused field can transfer momentum to the electron in principle along any direction. The four-momentum transfer $\Delta \pi_{\text{field}}^{\mu}(x_l)$ at each emission point $x_l$ from the field can be estimated from the relations $0=\Delta \pi^{\mu}(x_l)=p^{\mu}_f+\hbar k^{\mu}-p^{\mu}_i-\Delta \pi_{\text{field}}^{\mu}(x_l)$ and by employing the classical solution in Eqs. (\ref{u_+^1})-(\ref{u_-^1}). Finally, the focusing of the laser is expected to alter also the electron emission spectrum. This can be already anticipated by estimating the ``instantaneous'' classical cut-off emission frequency $\omega_c\sim \varepsilon^3/m^3\rho$, where $\rho$ is the curvature radius of the electron trajectory at the instant of emission \cite{Landau_b_2_1975}. By calculating $\rho$ from Eqs. (\ref{u_+^1})-(\ref{u_-^1}), one estimates $\omega_c\sim (|e|\varepsilon_0^2/m^5)\sqrt{(E_z+B_x)^2+(E_x-B_z)^2}$, with the second term inside the square root vanishing identically in the plane-wave case.

I am grateful to K. Z. Hatsagortsyan, D. V. Karlovets, C. H. Keitel, F. Mackenroth, S. Meuren, N. Neitz, M. Tamburini, and E. Yakaboylu for useful discussions.

\appendix

\section{Supplemental Material}

Based on the general argument that the de Broglie length of an ultrarelativistic particle is very small, here, we would like to solve the Dirac equation
\begin{equation}
[\gamma^{\mu}(i\hbar\partial_{\mu}-eA_{\mu})-m]\psi=0
\end{equation}
by applying the Wentzel-Kramers-Brillouin (WKB) method (the same notation is employed as in the main text) \cite{Landau_b_3_1977}. More precise conditions of validity of the present approach are provided below. 

According to the WKB method, we look for a solution of the form $\psi(x)=\exp[iS(x)/\hbar]\varphi(x)$ \cite{Pauli_1932,Rubinow_1963,Maslov_1981} and the Dirac equation becomes
\begin{equation}
\label{Dirac}
[\gamma^{\mu}(\partial_{\mu}S+eA_{\mu})+m]\varphi=i\hbar\gamma^{\mu}\partial_{\mu}\varphi.
\end{equation}
We first neglect the term proportional to $\hbar$ in this equation. The resulting equation
\begin{equation}
[\gamma^{\mu}(\partial_{\mu}S+eA_{\mu})+m]\varphi=0
\end{equation}
for the zero-order bi-spinor $\varphi(x)$ admits a non-trivial solution only if $\det[\gamma^{\mu}(\partial_{\mu}S+eA_{\mu})+m]=0$. As it is known \cite{Pauli_1932,Rubinow_1963}, this condition implies that $S(x)$ has to satisfy the Hamilton-Jacobi equation
\begin{equation}
(\partial_{\mu}S+eA_{\mu})(\partial^{\mu}S+eA^{\mu})-m^2=0
\end{equation}
and that it coincides with the classical action \cite{Landau_b_2_1975}. By applying the method of characteristics (see, e.g., \cite{Courant_b_1989}), the solution of the partial differential Hamilton-Jacobi equation is reduced to the solution of the ordinary differential Lorentz equation. Then, the action can be constructed from the equation \cite{Landau_b_2_1975}
\begin{equation}
\frac{dS}{ds}=-m-\frac{e}{m}p_{\mu}A^{\mu}, 
\end{equation}
where all quantities are calculated along the electron's trajectory/characteristics, parametrized via the proper time $s$. Since we have already determined the electron's trajectory within the required approximation (see Eqs. (6)-(8) in the main text), it is straightforward to integrate the equation for the action $S_{p_0}(\tau)$ by parametrizing the electron's trajectory via the variable $\tau$. After the integration has been carried out, one has to eliminate the initial quantities $\bm{r}_0$ in terms of the quantities $\bm{r}(\tau)=\bm{r}$ via Eq. (5) in the main text. The final expression of the action $S_{p_0}(\tau,\bm{r})$ calculated up to terms proportional to $1/p_{0,+}$ is
\begin{equation}
\label{S^1_SM}
\begin{split}
S^{(1)}_{p_0}(\tau,\bm{r})=&-p_{0,+}\phi-\frac{m^2}{2p_{0,+}}\tau\\
&-\int_{\tau_0}^{\tau}d\tau'\bigg[eA_-(\tau',\bm{r})+\frac{e^2}{2p_{0,+}}\bm{G}^2_p(\tau',\bm{r})\bigg].
\end{split}
\end{equation}

Now, the remaining task is to determine the bi-spinor $\varphi_{p_0,\sigma_0}(x)$, which solves the equation 
\begin{equation}
[\gamma^{\mu}p_{\mu}(x)-m]\varphi_{p_0,\sigma_0}=0,
\end{equation}
where the four-momentum $p^{\mu}(x)=-\partial^{\mu}S(x)-eA^{\mu}(x)$ has to be intended as a function of the coordinates like the action, and where the quantum number $\sigma_0=\pm 1$ indicates the electron's spin degree of freedom. By recalling the solution of the Dirac equation for a free electron \cite{Landau_b_4_1982}, we can already write $\varphi_{p_0,\sigma_0}(x)$ in general as
\begin{equation}
\label{phi_0}
\varphi_{p_0,\sigma_0}(x)=\frac{1}{\sqrt{2\varepsilon_0}}
\begin{pmatrix}
\sqrt{\varepsilon(x)+m}w_{p_0,\sigma_0}(x)\\
\frac{\bm{p}(x)\cdot\bm{\sigma}}{\sqrt{\varepsilon(x)+m}}w_{p_0,\sigma_0}(x)
\end{pmatrix},
\end{equation}
where $\bm{\sigma}$ are the Pauli matrices, $w_{p_0,\sigma_0}(x)$ is an arbitrary spinor, and where the normalization factor $1/\sqrt{2\varepsilon_0}$ has been introduced for convenience (a unity quantization volume is assumed). In order to determine the spinor $w_{p_0,\sigma_0}(x)$, we consider now Eq. (\ref{Dirac}) at first order in $\hbar$ and notice that it implies that 
\begin{equation}
[\gamma^{\mu}p_{\mu}(x)+m]\gamma^{\nu}\partial_{\nu}\varphi_{p_0,\sigma_0}=0. 
\end{equation}
By substituting the bi-spinor (\ref{phi_0}) in this equation, the latter reduces to the following equation for the spinor $w_{p_0,\sigma_0}(x)$ (see also \cite{Bagrov_1993}):
\begin{equation}
\label{Eq_w}
\begin{split}
p^{\mu}\partial_{\mu}w_{p_0,\sigma_0}=&-\frac{1}{2}(\partial_{\mu}p^{\mu})w_{p_0,\sigma_0}\\
&+\frac{ie}{2}\bm{\sigma}\cdot\left(\bm{B}-\frac{\bm{p}\times\bm{E}}{\varepsilon+m}\right)w_{p_0,\sigma_0}.
\end{split}
\end{equation}
This equation can also be solved via the method of characteristics by parametrizing the trajectory via the variable $\tau$. By setting 
\begin{equation}
\label{w_rho}
w_{p_0,\sigma_0}(\tau)=\exp\left[-\frac{1}{2}\int_{\tau_0}^{\tau}\frac{d\tau'}{p_+}(\partial_{\mu}p^{\mu})\right]\rho_{p_0,\sigma_0}(\tau),
\end{equation}
the spinors $\rho_{p_0,\sigma_0}(\tau)$ fulfill time-independent normalization conditions, which can be chosen as: $\rho^{\dag}_{p_0,\sigma_0}(\tau)\rho_{p_0,\sigma_0'}(\tau)=\delta_{\sigma_0,\sigma_0'}$. Also, the spin four-vector $\zeta^{\mu}$ in units of $\hbar$, defined as $\zeta^{\mu}=\bar{\psi}\gamma^5\gamma^{\mu}\psi/\bar{\psi}\psi=(\zeta^0,\bm{\zeta})$, with $\gamma^5=-i\gamma^0\gamma^1\gamma^2\gamma^3$ for a state $\psi(x)$ \cite{Landau_b_4_1982}, is given in our case by \cite{Landau_b_2_1975,Landau_b_4_1982}
\begin{equation}
\zeta^{\mu}(x)=\left(\frac{\bm{\zeta}_0\cdot\bm{p}(x)}{m},\bm{\zeta}_0+\frac{\bm{\zeta}_0\cdot\bm{p}(x)}{m[m+\varepsilon(x)]}\bm{p}(x)\right),
\end{equation}
with $\bm{\zeta}_0=\rho^{\dag}_{p_0,\sigma_0}\bm{\sigma}\rho_{p_0,\sigma_0}$. By evaluating the four-vector $\zeta^{\mu}(x)$ along the electron's trajectory, it can be shown that it satisfies the Bargmann-Michel-Telegdi equation 
\begin{equation}
\frac{d\zeta^{\mu}}{ds}=\frac{e}{m}F^{\mu\nu}\zeta_{\nu}
\end{equation}
for a Dirac electron with spin gyromagnetic factor equal to two \cite{Rubinow_1963}. By proceeding analogously as in the case of the action, one obtains the following first-order expression of $w_{p_0,\sigma_0}(\tau,\bm{r})$:
\begin{equation}
\begin{split}
w^{(1)}_{p_0,\sigma_0}(\tau,\bm{r})=&w_{0,p_0,\sigma_0}-\frac{e}{2}\int_{\tau_0}^{\tau}\frac{d\tau'}{p_{0,+}}\{\bm{\nabla}_{\perp}\cdot\bm{G}_p(\tau',\bm{r})\\
&-i\bm{\sigma}\cdot[\bm{B}(\tau',\bm{r})-\bm{n}\times\bm{E}(\tau',\bm{r})]\}w_{0,p_0,\sigma_0},
\end{split}
\end{equation}
where $w_{0,p_0,\sigma_0}=w_{p_0,\sigma_0}(\tau_0)$ is the initial spin vector, normalized as $w^{\dag}_{0,p_0,\sigma_0}w_{0,p_0,\sigma_0'}=\delta_{\sigma_0,\sigma_0'}$ (see discussion below Eq. (\ref{Eq_w})) and assumed to satisfy the equation $\bm{n}\cdot\bm{\sigma}w_{0,p_0,\sigma_0}=\sigma_0 w_{0,p_0,\sigma_0}$. Finally, by expanding the resulting state $\varphi_{p_0,\sigma_0}(\tau,\bm{r})$ up to terms proportional to $1/p_{0,+}$ in the expression $\psi_{p_0,\sigma_0}(\tau,\bm{r})=\exp[iS_{p_0}(\tau,\bm{r})/\hbar]\varphi_{p_0,\sigma_0}(\tau,\bm{r})$ of the electron state up to first order in $1/p_{0,+}$, we obtain:
\begin{equation}
\label{psi_0_SM}
\begin{split}
\psi^{(1)}_{p_0,\sigma_0}(\tau,\bm{r})=&e^{iS^{(1)}_{p_0}(\tau,\bm{r})/\hbar}\bigg[ 1-\frac{e}{2}\int_{\tau_0}^{\tau}\frac{d\tau'}{p_{0,+}}\{\bm{\nabla}_{\perp}\cdot\bm{G}_p(\tau',\bm{r})\\
&-i\bm{\Sigma}\cdot[\bm{B}(\tau',\bm{r})-\bm{n}\times\bm{E}(\tau',\bm{r})]\}\bigg] \frac{u_{p_0,\sigma_0}}{\sqrt{2\varepsilon_0}},
\end{split}
\end{equation}
where $\bm{\Sigma}=-i\gamma^1\gamma^2\gamma^3\bm{\gamma}$ and $u_{p_0,\sigma_0}$ is the usual constant free bi-spinor \cite{Landau_b_4_1982}. 

Note that up to first order in $1/p_{0,+}$ it is
\begin{equation}
\label{norm}
\psi^{(1)\,\dag}_{p_0,\sigma_0}(\tau,\bm{r})\psi^{(1)}_{p_0,\sigma_0}(\tau,\bm{r})\approx 1-e\int_{\tau_0}^{\tau}\frac{d\tau'}{p_{0,+}}\bm{\nabla}_{\perp}\cdot\bm{G}_p(\tau',\bm{r}).
\end{equation}
Thus, since the integral over the whole space of the quantity $\bm{\nabla}_{\perp}\cdot\bm{G}_p(\tau',\bm{r})$ vanishes, it is $\int d\bm{r}\psi^{(1)\,\dag}_{p_0,\sigma_0}(\tau,\bm{r})\psi^{(1)}_{p_0,\sigma_0}(\tau,\bm{r})=1+O(1/p^2_{0,+})$. Note that the same normalization condition holds also in the ordinary coordinate space $\bm{x}$, as the operator $\bm{\nabla}_{\perp}$ can be expressed as a linear combination of the derivatives with respect to the ordinary coordinates.

In order to determine the validity conditions of the approach presented above, we first request that
the first-order corrections in $1/p_{0,+}$ in the action and in the pre-factor matrix in the state
$\psi^{(1)}_{p_0,\sigma_0}(\tau,\bm{r})$ in Eq. (\ref{psi_0_SM}) are much smaller than the leading-order terms. Equations (\ref{S^1_SM}) and (\ref{psi_0_SM}) indicate that, as expected, such conditions are in order of magnitude equivalent to those obtained for the validity of the classical approximated solution (see discussion below Eq. (8) in the main text). More precise conditions also depend on the process to be investigated and on its formation region. Additional conditions for the validity of the above wave functions are obtained by estimating the size of the terms, which have been neglected in the original Dirac equation during the derivation. This can be carried out more easily by means of an alternative derivation of the wave function $\psi^{(1)}_{p_0,\sigma_0}(\tau,\bm{r})$ in Eq. (\ref{psi_0_SM}), starting from the ``squared'' Dirac equation \cite{Landau_b_4_1982}
\begin{equation}
\label{Dirac_s}
(i\hbar\partial_{\mu}-eA_{\mu})[(i\hbar\partial^{\mu}-eA^{\mu})\psi]-m^2\psi-\frac{ie\hbar}{2}\sigma^{\mu\nu}F_{\mu\nu}\psi=0.
\end{equation}
By writing again the wave function $\psi(x)$ in the form $\psi(x)=\exp[iS(x)/\hbar]\varphi(x)$ and by collecting the terms with the same power of $\hbar$, one obtains
\begin{equation}
\label{Dirac_s_2}
\begin{split}
&[(\partial_{\mu}S+eA_{\mu})(\partial^{\mu}S+eA^{\mu})-m^2]\varphi\\
&\quad-i\hbar\left[2(\partial_{\mu}S+eA_{\mu})\partial^{\mu}+\square S+\frac{e}{2}\sigma^{\mu\nu}F_{\mu\nu}\right]\varphi\\
&\quad\quad-\hbar^2\square\varphi=0,
\end{split}
\end{equation}
where $\square=\partial_{\mu}\partial^{\mu}$. At zero order in $\hbar$, we again obtain the Hamilton-Jacobi equation for the quantity $S(x)$, which we identify, as above, with the classical action. Thus, the four-vector $p^{\mu}(x)=-\partial^{\mu}S(x)-eA^{\mu}(x)$ is the corresponding classical electron four-momentum and, with this choice, Eq. (\ref{Dirac_s_2}) becomes
\begin{equation}
\label{eq_phi}
2p_{\mu}\partial^{\mu}\varphi+(\partial_\mu p^{\mu})\varphi-\frac{e}{2}\sigma^{\mu\nu}F_{\mu\nu}\varphi=-i\hbar\square\varphi.
\end{equation}
Now, our solution above corresponds to neglecting the term on the right-hand side proportional to $\hbar$. This is a good approximation if the formal operator conditions
\begin{align}
\hbar \left|\int_{\tau_0}^{\tau}\frac{d\tau'}{p_+}\square\right|\ll 1, && \hbar|\square|\ll |(\partial_\mu p^{\mu})|, && \hbar|\square|\ll |e|F_0
\end{align}
are fulfilled. We can ``estimate'' the  differential and integral operators starting from the spatio-temporal extension of the background field and by applying the operators to the bi-spinor $\varphi^{(1)}_{p_0,\sigma_0}(\tau,\bm{r})=\exp[-iS^{(1)}_{p_0}(\tau,\bm{r})/\hbar]\psi^{(1)}_{p_0,\sigma_0}(\tau,\bm{r})$ (see Eq. (\ref{psi_0_SM})). Also, we specialize to the case of a focused Gaussian laser field as that considered in the main text (central angular frequency $\omega_0$, central wavelength $\lambda_0=2\pi/\omega_0$, waist size $w_0$, Rayleigh length $l_R=\omega_0 w_0^2/2$, and pulse duration $T$). For the sake of simplicity and recalling that $c=1$ in our units, we also assume that $T\sim l_R$ (for experimental laser parameters of a tightly focused Ti:sapphire laser: $\omega_0=1.55\;\text{eV}$, $w_0=1\;\text{$\mu$m}$, and $T=10\;\text{fs}$, it is $l_R\approx 4.0\;\text{$\mu$m}$ and $T\approx 3.0\;\text{$\mu$m}$). Since $\square=2\partial^2/\partial\phi\partial\tau-\nabla^2_{\perp}$, an inspection to Eq. (\ref{psi_0_SM}) indicates that $|\square|\sim |e|F_0/w_0\varepsilon_0$. Analogously, it can be shown that $|(\partial_\mu p^{\mu})|\sim |e|F_0\lambda_0/w_0$, such that, being understood that the classical strong inequality $m\xi_0\ll\varepsilon_0$ is satisfied, the above conditions are fulfilled in order of magnitude if $\hbar/w_0\varepsilon_0\ll 1$. This condition is safely satisfied in the case of a tightly focused ($w_0\sim\lambda_0$), optical ($\lambda_0\sim 1\;\text{$\mu$m}$) laser field, as $\hbar/\varepsilon_0\sim m\lambda_C/\varepsilon_0\ll\lambda_C$, with $\lambda_C=\hbar/m=3.9\times 10^{-11}\;\text{cm}$ being the Compton wavelength.

\end{document}